\documentstyle[12pt]{article}

\makeatletter
\def\ps@myheadings{\let\@mkboth\@gobbletwo
 \def\@oddhead{\hfil\rightmark}%
 \def\@oddfoot{\hfil\rm\thepage\hfil}%
 \def\@evenhead{\@oddhead}%
 \def\@evenfoot{\@oddfoot}\def\sectionmark##1{}\def\subsectionmark##1{}}
\makeatother

\newcommand{\langleNS}{{}_{\rm NS}\langle}
\newcommand{\rangleNS}{\rangle_{\rm NS}}
\newcommand{\langleR}{{}_{\rm R}\langle}
\newcommand{\rangleR}{\rangle_{\rm R}}

\newcommand{\e}{{\rm e}}
\renewcommand{\d}{{\rm d}}

\newcommand{\mathrm}[1]{{\rm #1}}
\newenvironment{ack}{\section*{Acknowledgements}}{}

\begin{document}
\renewcommand{\rightmark}{\parbox{3in}{\flushright%
NBI-HE-97-05\\
NORDITA 97/7P\\
hep-th/9701190\\
January 1997
(Revised March 1997)
}}

\title{Boundary states for moving D-branes\thanks{This work is
    partially supported by the European Commission TMR programme
    ERBFMRX-CT96-0045 and by the INFN, Italy.}}

\author{
Marco Bill\'o\thanks{e-mail: billo@nbi.dk}\ \ and
Paolo Di Vecchia\thanks{e-mail: divecchia@nbi.dk}\\[6pt]
{\normalsize\sl NORDITA}\\
{\normalsize\sl Blegdamsvej 17, DK-2100 Copenhagen \O, Denmark}
\and
Daniel Cangemi\thanks{e-mail: cangemi@nbi.dk}\\[6pt]
{\normalsize\sl Niels Bohr Institute}\\
{\normalsize\sl Blegdamsvej 17, DK-2100 Copenhagen \O, Denmark}
}

\date{}

\maketitle
\thispagestyle{myheadings}
\begin{abstract}
  We determine the boundary state for both the NS-NS and R-R sectors of
  superstring theory. We show how they are modified under a boost. The boosted
  boundary state is then used for computing the interaction of two D-branes
  moving with constant velocity reproducing with a completely different method
  a recent calculation by Bachas.
\end{abstract}

\noindent
In the early years of string theory it was realized that the open
string planar loop could be factorized in the closed string channel in
which a closed string is disappearing in the
vacuum~\cite{LOVELACE,CS}. The boundary states were originally
introduced~\cite{GROUP} to write the planar loop for the oriented
string with gauge group SU$(N)$ in the following very simple
form\footnote{We use the conventions of Ref.~\cite{GSW} with
  $\eta_{\mu\nu} = \mathrm{diag} (-1, 1, \ldots, 1)$ and we denote $|
  {\hat{\lambda}}\rangle = (-1)^{\sum_{n=1}^{\infty} a^{\dagger}_{n}
    \cdot a_{n}} |\lambda \rangle$.}:
\begin{equation} 
  (2 \pi)^d \; \delta^{(d)} \! \left( \sum_{i=1}^{M} p_i\right) A_P (p_1
  \dots p_M ) = \frac{N}{(2 \alpha ')^{d/2}} \frac{ \langle  B_0 | D |
    B_M \rangle}{\pi}~.
  \label{planar}
\end{equation}
The boundary state with open strings attached on it is given by
\begin{equation}
  | B_M \rangle =  |B\rangle_{\mathrm{gh}}
  \sum_{\lambda, \mu} |\lambda \rangle_a | {\hat{\mu}}
  \rangle_{{\tilde{a}}} |p=0 \rangle 
  \frac{\langle  \mu , p=0 | T_M | \lambda , p=0 \rangle }{\langle  p=0
    | p=0\rangle}~, 
  \label{boundaM}
\end{equation}
where
\begin{eqnarray}
    T_M &=& \mathop{\mathrm{Tr}} \left( \lambda^{a_1} \dots
    \lambda^{a_M}\right)  
    \left[ 2 g_S \left(2 \alpha ' \right)^{\frac{d-2}{4}}  \right]^M
    \nonumber\\
  &&\times
  ( - 2 \pi \mathrm{i} )^M \int_{0}^{1} \d \nu_M
    \int_{0}^{\nu_M} \d \nu_{M-1} \dots 
  \int_{0}^{\nu_3} d \nu_2 \prod_{i=1}^{M} 
  \left(\frac{V_i (p_i, \e^{2 \pi \mathrm{i} \nu_i
    })}{\e^{2 \pi \mathrm{i} \nu_i }} \right)~.  
  \label{tm}
\end{eqnarray}
The vertex operators $V_i$ correspond to the emitted open string
states. The state $|p=0\rangle$ is the state with zero momentum and
$|\lambda \rangle$ and $|\mu\rangle$ are arbitrary states with norm
equal to one.  Finally $|B\rangle_{\mathrm{gh}}$ is the ghost part of
the boundary state that can be found in Ref.~\cite{CALLAN}.  When no
open string is attached to the boundary Eq.~(\ref{boundaM}) becomes
\begin{equation}
    \langle B_0 | = {}_{\mathrm{gh}}\langle B| \langle  p=0 |
    \sum_{\lambda} {}_{\tilde{a}}\langle \hat{\lambda} | 
    {}_a \langle \lambda |
  = {}_{\mathrm{gh}}\langle  B| \langle p=0 |\,\, {}_{\tilde{a}}
    \langle  0| {}_a \langle  0| \e^{  
   - \sum_{n=1}^{\infty} a_n \cdot {\tilde{a}}_n }~. 
  \label{bounda0}
\end{equation}
In the one loop case the ghost part has the only effect of eliminating
in the partition function two powers of the $26$ ones coming from the
contribution of the non-zero modes of the string coordinates and of
shifting by one the intercept of the Regge trajectory in the
propagator. Therefore for our purposes in this paper we can limit
ourselves to the string coordinates provided that we subtract two
powers of the $26$ ones coming from the non-zero modes of the string
coordinates and that we use the following closed string
propagator\footnote{When dealing with the superstring, we have $d=10$
  and the intercepts in the closed string propagators are $(1/2, 1/2)$
  for the NS-NS sector and $(0, 0)$ for the R-R sector.}:
\begin{equation}
  D = \frac{1}{2 \pi} \int_{|z| \leq 1} \frac{\d^2 z}{|z|^2} z^{L_0 -1} 
  {\bar{z}}^{{\tilde{L}}_0 -1 }~,
  \label{clopropa}
\end{equation}
with the integral being done on the unit disk. 

The formalism of the boundary state was subsequently and extensively
developed to include a slowly varying external gauge
field~\cite{CALLAN,CALLAN1}, the superstring~\cite{POLCHINSKI,CALLAN2}
and Dirichlet boundary conditions~\cite{GREEN94}.  It has been
recently used in Refs.~\cite{MIAO,CK,SCHMIDHUBER} to study the
properties of D-branes, but the authors of the above references have
mostly limited themselves to consider only the massless content of the
boundary state. It has also been discussed in the framework of the
light-cone quantization~\cite{GG} and of the D0-brane
dynamics~\cite{HIRANO,IENGO}.

It can be checked that the boundary state is a physical BRST invariant state.
This means that a closed string theory contains also these physical states in
addition to the BRST invariant perturbative string states. They can for
instance be used to compute open string multiloop diagrams by saturating the
$N$-string closed string vertex with $N$ of them. We will see later on that
these BRST invariant states can be generalized to satisfy also Dirichlet
boundary conditions and in this case they will be describing
D-branes~\cite{POLCHINSKI2}.

In this short note we explicitly construct the D-brane boundary state
for the bosonic string and for both the NS-NS and the R-R sectors of
the superstring.  We then boost these states with a Lorentz
transformation obtaining a boundary state describing a D-brane moving
with constant velocity. Finally we use these boosted states to compute
the vacuum energy obtaining precisely the expression previously found
by Bachas~\cite{BACHAS} using open strings. This alternative looks
promising and opens the way to more complicated situations, for
example the scattering of more than two branes, higher loop
corrections to the potential between two D-branes, D-branes
excitations, {\it etc.}

In the case of a D-brane the boundary state must be suitably modified.
The components longitudinal to the D $p$-brane (that we will sometimes
denote as $\|$) and the time component are treated as in the previous
case, and we will reserve the indices $\alpha$ and $\beta$ for them.
For the transverse components (denoted by $\bot$, or by the indices
$i,j$ and $k$) we must change the sign of the operators ${\tilde{a}}$,
corresponding to Dirichlet boundary conditions for the open string.
With these modifications the boundary state in Eq.~(\ref{bounda0})
becomes~\cite{CK}
\begin{equation} 
  |B, y \rangle = (2 \pi \sqrt{\alpha'})^{d_{\bot}}\,\delta^{(d_{\bot})}
  (q^\bot - y^\bot )\, \e^{ -
    \sum_{n=1}^{\infty} a_{n}^{\dagger} \cdot {\tilde{a}}_{n}^{\dagger}
  } | 0\rangle_a |0\rangle_{\tilde{a}} | p= 0\rangle~,
  \label{bounda}
\end{equation}
where the 
position $y$ of the D-brane is fixed by a $\delta$-function in $d_\bot = d -
p - 1$ dimensions.  We omit to write the boundary state for the ghost degrees
of freedom since its effect for our calculation will be exactly the same as in
the case of the planar loop. As already discussed the non-zero mode part in the
previous equation should be understood as follows:
\begin{equation}
  a_{n}^{\dagger} \cdot {\tilde{a}}_{n}^{\dagger} = 
  \eta_{\alpha\beta} a_n^{\dagger\alpha} a_n^{\dagger\beta} -
  \delta_{ij} a_n^{\dagger i} a_n^{\dagger j}~.
  \label{sign}
\end{equation}
The boundary state~(\ref{bounda}) satisfies the 
conditions~\cite{CK,SCHMIDHUBER,GG}
\begin{equation}
  \partial_{\tau} X^\alpha|_{\tau=0} \; |B,y\rangle =
  \partial_{\sigma} X^i |_{\tau=0} \; |B,y\rangle=0~.
\label{chara}
\end{equation}
In addition, it is crucial in our analysis to fix the location of the D-brane
by imposing an extra condition on the position operator in the transverse
direction corresponding to the $\delta$-function in Eq.~(\ref{bounda}). Thus,
the boundary state~(\ref{bounda}) is characterized by the conditions
\begin{equation}
  P^\alpha|_{\tau=0} \; |B ,y \rangle =0~, \qquad\qquad
  ( X^i - y^i)|_{\tau=0} \; |B ,y \rangle =0
  \label{extra}
\end{equation}
on its longitudinal momentum and transverse coordinates.
 
The D-brane can be boosted by applying to the boundary state the boost
operator
\begin{equation}
  |B, y , {\bf v} \rangle = \e^{\mathrm{i} v^j J^{0}{}_j} \;
  |B, {}^{({\bf v})} y \rangle~, 
  \label{boosted}
\end{equation}
where the ``velocity'' ${\bf v}$ is taken transverse to the D-brane, ${}^{({\bf
    v})}y^i = y^i + v^i (v_j y^j) (\cosh |{\bf v}| -1) / {\bf
    v}^2$ is the boosted position of the D-brane and
the generator of the Lorentz transformation is equal to
\begin{equation}
  J^{\mu \nu} = q^{\mu} p^{\nu} - q^{\nu} p^{\mu} - \mathrm{i}
  \sum_{n=1}^{\infty} 
  \left( a_{n}^{\dagger \mu } a_{n}^{ \nu } - a_{n}^{\dagger \nu } 
    a_{n}^{ \mu } +  {\tilde{a}}_{n}^{\dagger\mu} {\tilde{a}}_{n}^{ \nu} - 
    {\tilde{a}}_{n}^{\dagger \nu} {\tilde{a}}_{n}^{ \mu}\right)~.
  \label{lorgen}
\end{equation}
Performing the boost on the non-zero modes we find
\begin{equation}
  \e^{\mathrm{i} v^j J^{0}{}_{j}}  a_{n}^{\dagger} \cdot
  {\tilde{a}}_{n}^{\dagger}  
  \e^{- \mathrm{i} v^j J^{0}{}_{j}} =  a_n^{\dagger\|} \cdot 
  {\tilde{a}}_n^{\dagger\|} - 
  \sum_{k \neq j} a_n^{\dagger k}  {\tilde{a}}_n^{\dagger k}- 
  \left( \begin{array}{cc} 
      a_{n}^{\dagger 0} & a_{n}^{\dagger j} \end{array}\right)  M ({\bf v})^2 
  \left( \begin{array}{cc} {\tilde{a}}_{n}^{\dagger 0} \\ 
      {\tilde{a}}_{n}^{\dagger j} \end{array} \right) ~, 
  \label{nonzero}
\end{equation}
where
\begin{equation}
  M ({\bf v}) = \left(
    \begin{array}{cc} \cosh v & n^k \sinh v \\
      n^j \sinh v & \quad \delta^{jk} + n^j n^k (\cosh v - 1)
    \end{array} \right)~, 
  \label{lortra}
\end{equation}
with $v \equiv |{\bf v}|$ and $ n^j \equiv v^j/|{\bf v}|$.  The action
of the boost on the zero modes is also easily computed and one gets
\begin{equation}
  \e^{\mathrm{i} v^{j}(q^{0} p_{j} - q_{j} p^{0})}
  \e^{\mathrm{i} q^{\bot} \cdot Q^{\bot}} 
  \e^{- \mathrm{i} v^{j}(q^{0} p_{j} - q_{j} p^{0})}
  = \e^{\mathrm{i} Q_{j} \left\{q^j+ n^{j} q^{0} \sinh v
      + n^j (n \cdot q) \left( \cosh v -1 \right)  \right\} }~.
  \label{zeroboo}
\end{equation}
Remembering that the physical velocity $V$ of the brane is related to
$v$ through the relation $V = \tanh v$, the previous expression
becomes
\begin{equation}
  \e^{\mathrm{i} Q_{j} \left\{q^j+ n^{j} q^{0} 
      \sqrt{\frac{V^2}{1- V^2}} + n^j (n \cdot q) 
      \left[ \frac{1}{\sqrt{1- V^2}} -1 
      \right]   \right\} }~.
  \label{zeroboo2}
\end{equation}
We choose now to boost in one of the transverse directions, that we
call $i$, and we fix $n^i = -1$ and $n^j =0$ for $j \neq i$.  Using
Eq.~(\ref{zeroboo2}) the $\delta$-function in the boundary state gets
modified as follows:
\begin{equation}
  \delta^{(d^{\bot})} ( q^{\bot} - y^{\bot} ) \rightarrow \sqrt{1- V^2}\,
  \delta (q^i - q^0 V - y^i)\,\prod_{j \neq i} \delta( q^j - y^j
  )~. 
  \label{deltamod}
\end{equation}
Notice that the boost of the zero mode part induces a normalization
factor containing the physical velocity of the D-brane and a
modification of the $\delta$-function $\delta(q^i - y^i)$.  Thus the
boosted boundary state becomes
\begin{eqnarray}
  \lefteqn{
    |B, y, v\rangle = (2\pi\sqrt{\alpha^\prime})^{d_\bot}\,
    \sqrt{1- V^2}\, \delta(q^i - q^0 V - y^i ) \, \prod_{j \neq i} \delta ( q^j
    - y^j ) }
  \nonumber\\
  & \displaystyle \times
  \prod_{n=1}^{\infty} \e^{ - \left\{ a_n^{\dagger\|} \cdot 
      {\tilde{a}}_n^{\dagger\|} - 
      \sum_{j \neq i} a_n^{\dagger j}  {\tilde{a}}_n^{\dagger j}- 
      \left( \begin{array}{cc} 
          a_{n}^{\dagger 0} & a_{n}^{\dagger i} \end{array}\right)  M(v)^2 
      \left( \begin{array}{cc} {\tilde{a}}_{n}^{\dagger 0} \\ 
          {\tilde{a}}_{n}^{\dagger i} \end{array}\right)   \right\}}
    \; |p=0\rangle 
  |0\rangle_a |0\rangle_{{\tilde{a}}}~, 
  \label{bounsta}
\end{eqnarray}
where $M(v)$ is the matrix in Eq.~(\ref{lortra}) with the above choice
for $n^j$.  The non-zero mode structure of the vertex agrees with the
expression found in Ref.~\cite{CK}. The Born-Infeld factor
$\sqrt{1-V^2}$ is the Lorentz contraction factor taken out of the
correctly boosted $\delta$-function.  We will see
that this boosted $\delta$-function is essential to reproduce the
result of Ref.~\cite{BACHAS}.

The next step is to compute the matrix element $\langle B, y_1 , v_1 |
D |B, y_2 , v_2 \rangle$, where the closed string propagator is given
in Eq.~(\ref{clopropa}) and the ``velocities'' $v_1$ and $v_2$ are
both taken in the common direction $i$.  The contribution of the
non-zero modes and the ghosts is equal to
\begin{equation}
  \prod_{n=1}^{\infty} \left[ ( 1 - (z {\bar{z}})^n )^{-(d-2)} 
    \frac{ ( 1 - (z {\bar{z}})^n )^{2}}{[ 1 - \e^{2 \pi \epsilon} 
      (z {\bar{z}})^n ] [ 1 - \e^{- 2 \pi \epsilon} (z {\bar{z}})^n ]}
    \right]~, 
  \label{nonze}
\end{equation}
with $ \pi \epsilon \equiv |v_1 - v_2| $.  The contribution of the zero modes
is equal instead to
\begin{equation}
  (2\pi \sqrt{\alpha^\prime})^{2 d_\bot} \,
  \frac{V_p}{(2 \pi)^p} \frac{1}{(2 \pi)^{2 d_{\bot}}} \,
  \frac{\sqrt{1-V_{1}^{2}} \sqrt{1-V_{2}^{2}}}{| V_1 - V_2 |} \, 
  \e^{ \frac{b^2}{\alpha ' \log r^2}} \left[ - \frac{\alpha '}{4 \pi} \log r^2 
  \right]^{-\frac{d_{\bot}-1}{2}}~, 
  \label{zemo}
\end{equation}
with $r \equiv |z|$. We have chosen the normalization
\begin{equation}
  \langle  p=0 | \e^{\mathrm{i} q \cdot Q} | p=0\rangle =
  \delta^{(d)} ( Q)~,  
  \hspace{2cm}
  \delta^{(d)} ( 0) = \frac{V_d}{(2 \pi)^d}
  \label{del}
\end{equation}
and we have introduced $b^2 \equiv \sum_{j \not= i} (y_1^j - y_2^j)^2$.  It is
easy to check that
\begin{equation}
  \frac{\sqrt{1-V_{1}^{2}} \sqrt{1-V_{2}^{2}}}{|V_1 - V_2|}=
  \frac{1}{\sinh \pi \epsilon}~.
  \label{fa}
\end{equation}
In conclusion we get the following expression for the matrix element:
\begin{eqnarray}
  \lefteqn{
    \langle B, y_1 , v_1 | D |B, y_2 , v_2 \rangle =
    \frac{V_p}{(2 \pi)^p}  
    \frac{(\alpha ')^{ d_{\bot}} }{(\alpha ' /2)^{(d_{\bot} - 1)/2}} 
    \int_{0}^{1}  \frac{\d r}{r} r^{-2}
    \left[ \prod_{n=1}^{\infty} \left( 1- r^{2n}
    \right)\right]^{-(d-2)} 
    } \hspace{.4in} \nonumber\\
  & \displaystyle \times 
  \e^{ \frac{b^2}{\alpha ' \log r^2}}  \left[- \frac{1}{2 \pi} \log r^2 
  \right]^{-\frac{d_{\bot}-1}{2}} \frac{1}{\sinh \pi \epsilon}
  \prod_{n=1}^{\infty} \frac{ ( 1 - r^{2n} )^{2}}{[ 1 - \e^{2 \pi 
      \epsilon} r^{2n} ] [ 1 - \e^{- 2 \pi \epsilon} r^{2n}
    ]}~. \nonumber\\ 
  \label{fimatriele}
\end{eqnarray}
In order to get the correctly normalized vacuum energy $F$ one must
multiply Eq.~(\ref{fimatriele}) by a normalization factor:
\begin{equation}
  F = \frac{1}{\pi} \left( 2 \alpha ' \right)^{- d/2} 
  \langle B, y_1 , v_1 | D |B, y_2 , v_2 \rangle~.
  \label{corno}
\end{equation}
Notice that this normalization factor is the same as in Eq.~(\ref{planar})
and does not depend on the dimension of the D-branes.  After the
change of variable $r = \e^{-\pi \tau}$ we can write the vacuum energy
in the final form:
\begin{eqnarray}
    F &=& V_p \left( 8 \pi^2 \alpha ' \right)^{-p/2} 
    \int_{0}^{\infty}  \d \tau  \, \tau^{-(d-2)/2 + p/2} 
    \, \e^{- \frac{b^2}{2 \pi \alpha ' \tau }}
    \e^{2 \pi \tau} \left[ \prod_{n=1}^{\infty} 
      \left( 1- \e^{- 2 \pi \tau n} \right)\right]^{-d+2}
    \nonumber\\
  && \times
  \frac{1}{2 \sinh \pi \epsilon}
  \prod_{n=1}^{\infty} \frac{ ( 1 - \e^{- 2\pi \tau n} )^{2}}{[ 1 - \e^{2 \pi 
      \epsilon} \e^{- 2\pi \tau n} ] [ 1 - \e^{- 2 \pi \epsilon} 
    \e^{-2 \pi \tau n} ]}~, 
  \label{fivacuum}
\end{eqnarray}
that reproduces Bachas' calculation for the bosonic string~\cite{BACHAS}.
Remember that in this case $d=26$.

Let us consider now the fermionic string.  The fermionic part of the boundary
states must obey the boundary conditions
\begin{eqnarray}
  \label{nb1}
  (\psi^\alpha - \mathrm{i} \eta {\tilde\psi}^\alpha)|_{\tau=0} \;
  |{\cal B},\eta\rangle & = & 0~, \nonumber \\
  (\psi^i + \mathrm{i} \eta {\tilde\psi}^i)|_{\tau=0} \; |{\cal
  B},\eta\rangle & = & 0~,
\end{eqnarray}
where $\eta=\pm 1$. By consistency, the fields $\psi^\mu$ and
${\tilde\psi}^\mu$ must have the same periodicity, and therefore only
the NS-NS and R-R sectors are to be considered.

In the NS-NS case, Eqs.~(\ref{nb1}) are solved by
\begin{equation}
  \label{nb2}
  |{\cal B},\eta\rangleNS = 
  \e^{\mathrm{i} \eta \sum_{n=1}^{\infty}   b_{n-1/2}^{\dagger} \cdot 
    {\tilde{b}}_{n-1/2}^{\dagger}} \; |0\rangle_b |0\rangle_{{\tilde{b}}}\, 
  |{\cal B}\rangle_{\mathrm{gh}}~.
  \label{ferboun}
\end{equation}
The dot in the exponent should be understood as in Eq.~(\ref{sign});
we do not write the ghost contribution that is explicitly given in
Refs.~\cite{POLCHINSKI,CALLAN2}.  It is easy to check that the
corresponding GSO projected state is
\begin{equation}
  \label{nb3}
  |{\cal B}\rangleNS \equiv {1 - (-)^{\tilde F} \over 2}\,
  {1 - (-)^F \over 2}\, |{\cal B},+\rangleNS = {1\over 2}
  \left(|{\cal B},+\rangleNS - |{\cal B},-\rangleNS\right)~,
\end{equation}
with $F$ being the fermion number operator: $F=\sum_{n=1}^\infty
b_{n-1/2}^\dagger\cdot b_{n-1/2}$.

In the R-R sector, we have 
\begin{equation}
  \label{nb4}
  |{\cal B},\eta\rangleR = 
  \e^{\mathrm{i} \eta \sum_{n=1}^{\infty}   d_n^{\dagger} \cdot 
    \tilde{d}_n^{\dagger}} \; |0\rangle_d |0\rangle_{\tilde{d}}\, 
  |\eta\rangleR \, |{\cal B}\rangle_{\mathrm{gh}}~,
  \label{ferbounr}
\end{equation}
where $|\eta\rangleR$ is the fermionic zero mode boundary state.  The
correct normalization of the R-R contribution to the boundary state
requires some care in the definition of $|\eta\rangleR$. The Clifford
algebras formed by the $d_0^\mu$'s and the one formed by the $\tilde
d_0^\mu$'s are conveniently represented by gamma matrices acting on
normalized sixteen-dimensional spinor states $|a\rangle$ and $|\tilde
a\rangle$ [$a, \tilde a = 1, \ldots, 16$]. Since moreover $d_0^\mu$
and $\tilde d_0^\nu$ anticommute, we need to introduce an extra
normalized two-dimensional spinor state $|m\rangle$ [$m=1,2$] on which
$d_0^\mu$ and $\tilde d_0^\nu$ act as Pauli matrices:
\begin{eqnarray}
  d_0^\mu \; |a\rangle |\tilde a\rangle |m\rangle &=& \frac{1}{\sqrt{2}}
  \; \sigma_1{}^m{}_n 
  \Gamma^\mu{}^a{}_b \; |b\rangle |\tilde a\rangle |n\rangle~, \\
  \tilde d_0^\mu \; |a\rangle |\tilde a\rangle |m\rangle &=&
  \frac{1}{\sqrt{2}} \; \sigma_2{}^m{}_n
  \Gamma^\mu{}^{\tilde b}{}_{\tilde a} \; |a\rangle |\tilde b\rangle
  |n\rangle~.
\end{eqnarray}
The Ramond zero mode state is then defined in terms of the $(p+1)$
chiral matrix $\Gamma_{p+2} = - \mathrm{i}^{p(p+1)/2 + 1} \Gamma^0 \Gamma^1
\cdots \Gamma^p$
(satisfying $\Gamma_{p+2}^2=1$ and $\Gamma_{p+2}^\dagger = 
\Gamma_{p+2}$)~\cite{GG,BG}:
\begin{eqnarray}
  |+\rangleR &=&\frac{\mathrm{i}}{2} \left( 1 + (-)^p \sigma_3 \right)^m{}_n
  \left( \Gamma_{p+2} \right)^{\tilde a}{}_a \; |a\rangle |\tilde
  a\rangle |n\rangle~, \\ 
  |-\rangleR &=& \frac{\mathrm{i}}{2} \left( 1 + (-)^p \sigma_3 \right)^m{}_n
  \left( \Gamma_{p+2}\Gamma_{11} \right)^{\tilde a}{}_a \; |a\rangle
  |\tilde a\rangle |n\rangle~,
\end{eqnarray}
with $m=1$ for $p$ even and $m=2$ for $p$ odd.  It obeys the zero mode
part of the conditions~(\ref{nb1}) and verifies the following
properties:
\begin{eqnarray}
  && 32 \, d_0^{11} \; |\eta\rangleR \equiv 32 \, d_0^0 d_0^1 \cdots d_0^9 \;
  |\eta\rangleR = |-\eta\rangleR = 
  (-)^p \, 32 \, {\tilde d}_0^{11} \; |\eta\rangleR~, \\
  && \langleR \eta | \eta' \rangleR = - 16 \; \delta_{\eta\eta'}~.
  \label{nb8}
\end{eqnarray} 
The crucial minus sign in the inner product comes from the exchange in
the ordering of the spinor states in the conjugate state.  The fermion
number operator in the R sector has a zero mode part: $(-)^F = 32 \,
d_0^{11} (-)^{\sum d^\dagger_n d_n}$, and analogously for $(-)^{\tilde
  F}$. In the above explicit representation, the GSO projected R-R
state becomes
\begin{equation}
  \label{nb9}
  |{\cal B}\rangleR \equiv {1 +(-)^p (-)^{\tilde F} \over 2}\,
  {1 + (-)^F \over 2}\, |{\cal B},+\rangleR = {1\over 2}
  \left(|{\cal B},+\rangleR + |{\cal B},-\rangleR\right)
\end{equation}
with the sign $(-)^p$ taken in order to get a non-vanishing result.
The GSO projection is thus of type IIB for $p$ odd, and of type IIA
for $p$ even, in accordance with the Ramond-Ramond charge that is
carried by the Dirichlet $p$-brane~\cite{POLCHINSKI3}.

As in the case of the bosonic oscillators we can boost the fermionic boundary
states with a Lorentz transformation:
\begin{equation}
  |{\cal B}, \eta, v_i \rangle = \e^{\mathrm{i} v^i
  J^{0}{}_{i}} \; |{\cal B}, \eta 
  \rangle~,  
  \label{booferm}
\end{equation}
where the generator of the Lorentz transformation is given by
\begin{equation}
  J^{\mu \nu} = - \mathrm{i} \sum_{n=1}^{\infty}
  \left( b_{n-1/2}^{\dagger \mu } b_{n-1/2 }^{ \nu } - b_{n-1/2 }^{\dagger \nu
  } 
    b_{n-1/2}^{ \mu } +  
    {\tilde{b}}_{n-1/2}^{\dagger \mu} {\tilde{b}}_{n- 1/2}^{ \nu} - 
    {\tilde{b}}_{n-1/2}^{\dagger \nu} {\tilde{b}}_{n- 1/2}^{
  \mu}\right) 
  \label{ferlorgen}
\end{equation}
for the Neveu-Schwarz sector and by
\begin{equation}
  J^{\mu \nu} = -\frac{\mathrm{i}}{2} [ d_{0}^{\mu}, d_{0}^{\nu} ]  
  -\frac{\mathrm{i}}{2} [ {\tilde{d}}_{0}^{\mu}, {\tilde{d}}_{0}^{\nu} ]
  - \mathrm{i} \sum_{n=1}^{\infty}
  \left( d_{n}^{\dagger \mu } d_{n}^{ \nu } - d_{n}^{\dagger \nu } 
    d_{n}^{ \mu } +  
    {\tilde{d}}_{n}^{\dagger \mu} {\tilde{d}}_{n}^{ \nu} - 
    {\tilde{d}}_{n}^{\dagger \nu} {\tilde{d}}_{n}^{ \mu}\right)
  \label{ferlorgenr}
\end{equation}
for the Ramond sector.  For the part containing only non zero modes we obtain
for both sectors an expression very similar to the one obtained in
Eq.~(\ref{nonzero}) for the non-zero mode of the bosonic coordinates:
\begin{eqnarray}
  \e^{\mathrm{i} v^i J^{0}{}_{i}}  b_{r}^{\dagger} \cdot
  {\tilde{b}}_{r}^{\dagger}  
  \e^{- \mathrm{i} v^i J^{0}{}_{i}} =  
  b_r^{\dagger ;\|} \cdot {\tilde{b}}_r^{\dagger ;\|}
  - \sum_{j \neq i} b_{r ;j}^{\dagger}  {\tilde{b}}_{r;j}^{\dagger}- 
  \left( \begin{array}{cc} 
      b_{r}^{0\dagger} & b_{r}^{i\dagger} \end{array}\right)  M(v)^2 \left( 
    \begin{array}{cc} {\tilde{b}}_{r}^{0\dagger} \\ 
      {\tilde{b}}_{r}^{i\dagger} \end{array} \right)~, 
  \label{fernonzero}
\end{eqnarray}
where $r$ is a positive half-integer for the Neveu-Schwarz sector. The
same holds true with $b_r$ replaced by $d_n$ [$n$ a positive integer]
for the Ramond sector.  The zero mode part of the boost operator is
conveniently expressed as
\begin{equation}
  \e^{\mathrm{i} v^i J^{0}{}_{i}} = \e^{v^{i} \left[
      d^{0 \, (+)} d_{i}^{(-)} +  
      d^{0 \, (-)} d_{i}^{(+)} \right]}
  \label{zemoboost}
\end{equation}
in terms of the combinations $d^{\mu \, (\pm) } =
\frac{1}{\sqrt{2}}\left( d_{0}^{\mu} \pm \mathrm{i}
  {\tilde{d}}_{0}^{\mu} \right)$, whose only non-zero anticommutators
are $\{ d^{\mu \, (+)} , d^{\nu \, (-)} \} = \eta^{\mu \nu}$. This
yields
\begin{equation}
  |\eta,v\rangleR \equiv
  \e^{\mathrm{i} v^i J^{0\,}_{i}} | \eta \rangleR = \left[
  \cosh v + \sinh v\, 
    d^{0 \,(+\eta)} d_{i}^{(-\eta )} \right] | \eta \rangleR
  \label{boozemo}
\end{equation}
and
\begin{equation}
  \langleR \eta , v_2 | \eta,  v_1 \rangleR  = -16 \cosh ( v_1 - v_2 )
  = - 16 \cosh \pi \epsilon~,
  \label{zemocontr}
\end{equation}
where we have taken into account that $d_0^{\alpha(-\eta)} |\eta\rangleR =
d_0^{i(\eta)} |\eta\rangleR = 0$ and the normalization in Eq.~(\ref{nb8}).

We now compute the matrix element of the closed string propagator
between two boundary states. With our conventions, the conjugate of a
fermionic boundary state defined as in Eq.~(\ref{nb1}) satisfies $0 =
\langle {\cal B},\eta| \; (\psi^\alpha + \mathrm{i} \eta
{\tilde\psi}^\alpha)|_{\tau=0} = \langle {\cal B},\eta| \; (\psi^i -
\mathrm{i} \eta {\tilde\psi}^i)|_{\tau=0}$. 

For the NS-NS part, we get
\begin{equation}
  \langleNS  {\cal B}, v_1| z^{L_0 - 1/2} {\bar{z}}^{{\bar{L}}_0 - 1/2} 
  |{\cal B}, v_2\rangleNS = \frac{1}{2r} \left[Z_{+} (\epsilon)
    - Z_{-} (\epsilon) \right]~,
  \label{fermatele}
\end{equation}
where the functions
\begin{equation}
  Z_{\pm} (\epsilon) =  \prod_{n=1}^{\infty} \left[ 1 \pm  r^{2n-1}
  \right]^8  
  \prod_{n=1}^{\infty} \frac{\left[1 \pm \e^{2 \pi \epsilon}
  r^{2n-1} \right] 
    \left[1 \pm \e^{-2 \pi \epsilon} r^{2n-1} \right]}{(1 \pm
  r^{2n-1})^2}~. 
  \label{fercontri}
\end{equation}
are related to the Jacobi's theta functions.

The R-R part reads
\begin{eqnarray}
  \lefteqn{
    \langleR  {\cal B}, v_1| z^{L_0} {\bar{z}}^{{\bar{L}}_0} 
    |{\cal B}, v_2\rangleR = -8\,\cosh \pi \epsilon
    }\hspace{1in}\nonumber\\
  &\displaystyle \times 
  \prod_{n=1}^{\infty} \left[ 1 +  r^{2n} \right]^8
  \prod_{n=1}^{\infty} \frac{\left[1 + \e^{2 \pi \epsilon}
    r^{2n} \right] 
    \left[1 + \e^{-2 \pi \epsilon} r^{2n} \right]}{(1 + r^{2n})^2}~,
  \label{fermateler}
\end{eqnarray}
where the factor $-8 \cosh \pi \epsilon$ comes from the boost of the zero modes
and the infinite product comes from the non-zero modes.

The sum of the NS-NS contribution, Eq.~(\ref{fermatele}), and of the
R-R contribution, Eq.~(\ref{fermateler}), can be recast in the
following form:
\begin{equation}
  \frac{1}{2r} \prod_{n=1}^{\infty} \left( 1 - r^{2n} \right)^{-4}
  \sum_{\alpha=2,3,4} e_{\alpha} \Theta_{\alpha} ( - \mathrm{i}
  \epsilon | \mathrm{i} \tau) 
  \left[ \Theta_{\alpha} ( 0 | \mathrm{i} \tau) \right]^3~, 
\end{equation}
with $ e_3 = 1 = - e_2 = - e_4$.

If we also include the bosonic contribution with the same
normalization as in the bosonic case, we arrive at the final result
for the vacuum energy $F$ in the case of superstring:
\begin{eqnarray}
    F &=& V_p \left( 8 \pi^2 \alpha ' \right)^{-p/2} 
    \int_{0}^{\infty}  \d \tau  \, \tau^{- 4 + p/2} 
    \, \e^{- \frac{b^2}{2 \pi \alpha ' \tau }} 
    \frac{1}{2} \sum_{\alpha=2,3,4} e_{\alpha} \Theta_{\alpha} ( -
    \mathrm{i} \epsilon | \mathrm{i}  
    \tau) \left[ \Theta_{\alpha} ( 0 | \mathrm{i} \tau) \right]^3
    \nonumber\\
  && \times
  \left[ \e^{- \pi \tau/12} \prod_{n=1}^{\infty} 
    \left( 1- \e^{- 2 \pi \tau n} \right)\right]^{-12}
  \frac{1}{2 \pi \mathrm{i}} 
  \frac{\Theta_1 ' ( 0 | \mathrm{i} \tau)}{\Theta_1
    ( - \mathrm{i} \epsilon | \mathrm{i} \tau)}~,
  \label{fivasu}
\end{eqnarray}
that exactly reproduces Bachas' calculation: $F = -
\delta_\mathrm{Bachas}$~\cite{BACHAS}.  An alternative form is~\cite{GG}
\begin{eqnarray}
  F &=& V_p \left( 8 \pi^2 \alpha ' \right)^{-p/2} 
  \int_{0}^{\infty}  \d \tau  \, \tau^{- 4 + p/2} 
  \, \e^{- \frac{b^2}{2 \pi \alpha ' \tau }} 
  \nonumber\\
  && \times
  \left[ \e^{- \pi \tau/12} \prod_{n=1}^{\infty} 
    \left( 1- \e^{- 2 \pi \tau n} \right)\right]^{-9}
  \frac{1}{\mathrm{i}}
  \frac{\left[\Theta_1 ( -\mathrm{i} \epsilon/2 | \mathrm{i} \tau)\right]^4}
       {\Theta_1 ( - \mathrm{i} \epsilon | \mathrm{i} \tau)}~.
  \label{fivasu2}
\end{eqnarray}
To compute the vacuum energy of a D-brane anti-D-brane system amounts 
simply to change the sign of the R-R contribution~\cite{BS}.

In this paper we have shown how the formalism based on the boundary
states considerably simplifies the calculation of the interaction
between two D-branes moving with constant velocity.  The boundary
state can also be extended to include an arbitrary number of both open
and closed strings attached to it and in fact constitutes the basic
element for computing the interaction between any number of D-branes
and closed and open strings in pretty much the same way as one can
compute multiloops in open string theory starting from the
multi-string vertex for closed strings and saturating the external
legs with boundary states describing the closed-string vacuum
transition after the insertion of a closed string propagator. Work is
in progress along these lines.\footnote{Actually the scattering of
  closed strings from many D-branes has been recently computed in
  Ref.~\cite{FRAU}.}

\begin{ack}
  DC is grateful to the Centre de Physique Th\'eorique of the Ecole
  Polytechnique in Paris for its hospitality while part of this work
  was completed and thanks C.~Bachas for helpful discussions.
\end{ack}


\begin{thebibliography}{99}
 
\bibitem{LOVELACE} C. Lovelace, Phys. Lett. 34B (1971) 500.
  
\bibitem{CS} L. Clavelli and J. Shapiro, Nucl. Phys. B 57 (1973) 490.
               
\bibitem{GROUP} M. Ademollo, A.~D'Adda, R. D'Auria, F. Gliozzi, E.
  Napolitano, S. Sciuto and P. Di Vecchia, Nucl. Phys. B 94 (1975)
  221.
  
\bibitem{GSW} M.B. Green, J.H. Schwarz and E. Witten, Superstring
  Theory (Cambridge University Press, New York, 1987).
  
\bibitem{CALLAN} C.G. Callan, C. Lovelace, C.R. Nappi and S.A. Yost,
  Nucl.  Phys. B 293 (1987) 83.
  
\bibitem{CALLAN1} C.G. Callan, C. Lovelace, C.R. Nappi and S.A. Yost,
  Nucl. Phys. B 288 (1987) 525.
  
\bibitem{POLCHINSKI} J. Polchinski and Y. Cai, Nucl. Phys. B 296
  (1988) 91.
  
\bibitem{CALLAN2} C.G. Callan, C. Lovelace, C.R. Nappi and S.A. Yost,
  Nucl. Phys. B 308 (1988) 221.
  
\bibitem{GREEN94} M.B. Green, Phys. Lett. B 329 (1994) 435; M.B. Green
  and P. Wai, Nucl. Phys. B 431 (1994) 131.
  
\bibitem{MIAO} Miao Li, Nucl. Phys. B 460 (1996) 351.
  
\bibitem{CK} C.G. Callan and I.R. Klebanov, Nucl. Phys. B 465 (1996)
  473.
  
\bibitem{SCHMIDHUBER} C. Schmidh{\"{u}}ber, Nucl. Phys. B 467 (1996)
  146.
  
\bibitem{GG} M.B. Green and M. Gutperle, Nucl. Phys. B 476 (1996) 484.
  
\bibitem{HIRANO} S. Hirano and Y. Kazama, 
  preprint UT-Komaba 96-26 (1996) hep-th/9612064.
    
\bibitem{IENGO} F.~Hussain, R.~Iengo and C.~N\'u\~nez, 
  preprint SISSA 3/97/EP (1997) hep-th/9701143.
 
\bibitem{POLCHINSKI2} J.~Polchinski, S.~Chaudhuri, C.V.~Johnson, Notes
  on D-branes, preprint NSF-ITP-96-003 (1996) hep-th/9602052.
  
\bibitem{BACHAS} C. Bachas, Phys. Lett. B 374 (1996) 49.
  
\bibitem{BG} O. Bergman and M.R. Gaberdiel, 
  preprint HUTP-97/A003 (1997) hep-th/9701137.
  
\bibitem{POLCHINSKI3} J. Polchinski, Phys. Rev. Lett. 75 (1995) 4724.
  
\bibitem{BS} T.~Banks and L.~Susskind, 
  preprint RU-95-87 (1995) hep-th/9511194.
    
\bibitem{FRAU} M. Frau, A. Lerda, I. Pesando, R. Russo and S.  Sciuto,
  preprint DFTT 8/97 (1997) hep-th/9702037.
 
\end{thebibliography}
\end{document}